\documentclass{chet}

\usepackage{graphicx}
\usepackage[crop=pdfcrop]{pstool}
\usepackage{dsfont}

\DeclareMathOperator{\Tr}{Tr}

\preprint{UCSD-PTH-11-17} 

\title{Cyclic unparticle physics} 

\author{Jean-Fran\c{c}ois Fortin, Benjam\'\i{}n Grinstein and Andreas
Stergiou\emails{jffortin@physics.ucsd.edu, bgrinstein@ucsd.edu,
stergiou@physics.ucsd.edu}}

\affiliation{Department of Physics, University of California, San Diego, La
Jolla, CA 92093 USA}

\abstract{We explore phenomenological consequences of coupling a
non-conformal scale-invariant theory to the standard model. We point out
that, under certain circumstances, non-conformal scale-invariant theories
have oscillating correlation functions which can dramatically modify
standard model processes.  We dub this scenario cyclic unparticle physics,
or simply cyclunparticle physics. We compute phase spaces and amplitudes
associated with final state cyclunparticle and cyclunparticle exchange,
respectively. We show detailed formulae in a simple example.}

\newcommand{\CQ}{\ensuremath{\mathcal{Q}}}
\newcommand{\CP}{\ensuremath{\mathcal{P}}}
\newcommand{\CO}{\ensuremath{\mathcal{O}}}

\newcommand{\vev}[1]{\langle{#1}\rangle}

\date{October 2011}

\begin{document}

\maketitle

\newsec{Introduction}
The physical consequences of a quantum field theory can all be obtained
from its $S$-matrix, provided the $S$-matrix is well defined.  Of course,
the $S$-matrix directly encodes information on scattering cross sections
and hence on widths and masses of resonances too. The computation of some
other quantities, such as static properties of stable particles like the
gyromagnetic ratio of the electron or the neutron, is only slightly less
direct. But even quantities such as the thermodynamic potential of a system
in a thermal bath, something that would appear quite unrelated to the
$S$-matrix, can be obtained form it
\rcite{Dashen:1969ep,Dashen:1974jw,Dashen:1974yy}.  However, the $S$-matrix
does not exist in theories with scale invariance.

In the absence of an $S$-matrix one is left to study correlation functions.
While these are not physical, they surely encode measurable information. To
address the question of how to access this information, Georgi proposed a
model with two sectors that are very weakly coupled through irrelevant
operators\rcite{Georgi:2007ek}. The first sector, call it the `SM' sector,
is by itself non-scale-invariant. An example of such a sector is the
standard model of electroweak and strong interactions. The second sector,
call it the `SI' sector, is, considered in its own right, a scale-invariant
theory (SIT).  In this construction an observer made up of SM stuff can
conduct experiments that excite and probe the SI sector. For one example,
scattering of SM particles into SI stuff appears like the scattering into
invisible particles with possibly fractional dimensions of phase space. And
for another, exclusive scattering processes of SM to SM particles can
present unusual patterns of SM-SI interference\rcite{Georgi:2007si}.
Georgi dubbed this construction `unparticle physics' because of the
apparent fractional dimension of the phase space in `unparticle'
production. 
 
In this letter we distinguish an SI sector that is merely scale-invariant
from one that displays full fledged conformal invariance. Virtually all of
the (too numerous to cite) literature on phenomenology of unparticles
assumes the SI sector is a conformal field theory (CFT). It was pointed out
in Ref.~\rcite{Grinstein:2008qk} that unitarity imposes constraints on the
dimensions of non-scalar unparticle operators that render them
phenomenologically less accessible. In avoiding these constraints, work on
unparticle phenomenology that appeared after Ref.~\rcite{Grinstein:2008qk}
often appeals to an SI sector that is not a CFT.\foot{Even if only scale
invariance is assumed, there are still constraints on dimensions of
operators; see footnote 9 in Ref.~\rcite{Grinstein:2008qk}.} But until
recently the structure of non-CFT SI systems was largely unknown. Our work
has uncovered a remarkable feature of non-CFT SI systems: rather than
corresponding to fixed points of an RG flow, they are found as limit cycles
(or ergodic trajectories) in the flows\rcite{Fortin:2011ks,Fortin:2011sz}.
This most remarkable feature of non-CFT scale-invariant theories has been
missed in all of the unparticle literature. We propose to investigate it.

This serves another purpose. One may wonder if these cycles, being
unfamiliar, have any physical consequences at all. One may suspect that
they don't, because on the SI cycle one may recast the RGE as one with a
vanishing beta function but with a universal shift of anomalous dimensions.
We will lay this question to rest once and for all: probing a cycle of the
SI system through the unparticle setup results in observable oscillations
as a function of energy in, for example, cross sections. In our mind,
settling this issue, which has been raised informally by many, is no less
important than phenomenological implications. 

We hasten to indicate that cyclic unparticles, or `cyclunparticles,' are
more conjectural than plain unparticles. For both, one conjectures the
existence of an extension to the SM Lagrangian consisting of an SI sector
and of irrelevant operators coupling the SM to the SI. But while the
existence of CFTs that can make up the SI sector in plain vanilla
unparticle scenarios is well established, at present we have only
established non-CFT scale-invariant flows in $D=4-\epsilon$ spacetime
dimensions. We must note that we have found exact solutions to the two-loop
beta functions that correspond to flows with limit cycles in $D=4$
spacetime dimensions. These solutions, however, could be destabilized by
three-loop effects which have not been studied. Our approach here is to
assume the existence of SITs even at strong coupling and use the properties
that follow from general considerations.

\newsec{The main ingredient}
The SI sector consists of $N\ge2$ real scalars, $n\ge1$ Weyl spinors and
non-Abelian gauge interactions. The kinetic energy terms for the real
scalars and Weyl fermions display $SO(N)$ and $U(n)$ symmetries,
respectively, which are broken by the scalars' quartic self-interactions
with coupling constants $\lambda_{abcd}$, the Yukawa interactions with
couplings $y_{a|ij}$, and by gauge interactions. Here $a,b,c,d$ run from 1
through $N$, while $i,j$ run from $1$ through $n$. At a scale but not
conformal invariant point there are solutions to
\twoseqn{ \beta_{abcd}-\CQ_{abcd}&=0,}[ScInvQ] {
\beta_{a|ij}-\CP_{a|ij}&=0.}[ScInvP][ScInvII] 
where
\twoseqn{ \CQ_{abcd}&=-Q_{a'a}\lambda_{a'bcd}+\text{3
permutations},}[QDefd] {
\CP_{a|ij}&=-Q_{a'a}y_{a'|ij}-P_{i'i}y_{a|i'j}-P_{j'j}y_{a|ij'}.}[PDefd][QPDefd]
such that the beta functions themselves do not all vanish
\rcite{Polchinski:1987dy, Dorigoni:2009ra}. That is, one can find a value
of the coupling constants {\it and} a real anti-symmetric matrix $Q$ and an
anti-Hermitian matrix $P$ such that Eqs.~\ScInvII are satisfied without
setting all terms to zero.

Once a solution has been found, the SI cycle is given by constant gauge
couplings and, with $t=\ln(\mu_0/\mu)$,
\twoseqn{
\lambda_{abcd}(t)&=\widehat{Z}_{a'a}(t)\widehat{Z}_{b'b}(t)\widehat{Z}_{c'c}(t)\widehat{Z}_{d'd}(t)\lambda_{a'b'c'd'}
, }[cycLamda] {
y_{a|ij}(t)&=\widehat{Z}_{a'a}(t)\widehat{Z}_{i'i}(t)\widehat{Z}_{j'j}(t)y_{a'|i'j'},
}[cycY][cycCouplings]
where $\widehat{Z}_{ab}(t)=(e^{Qt})_{ab}$ and
$\widehat{Z}_{ij}(t)=(e^{Pt})_{ij}$ are clearly elements of $SO(N)$ and
$U(n)$, respectively. 

Let us remark here that scale-invariance \emph{requires} that $Q$ and $P$
be constant. To see that, notice that once a solution to Eqs.\ \ScInvII is
found, then Eqs.\ \ScInvII remain satisfied even if the running couplings
\cycCouplings are used. Therefore, if one finds a scale-invariant point,
one can be sure that there is a trajectory going through that point with
the same $Q$ and $P$. Trajectories with $Q$ and/or $P$ that are functions
of RG time are not possible, for then those, at every value of $Q$ and $P$,
would intersect trajectories with constant $Q$ and $P$.

For simplicity we will assume that $P=0$ in what follows. The interesting
effects arise from orbits in either of the groups $SO(N)$ and $U(n)$, so
considering only one will suffice to illustrate the main features of this
novel physics.

We assume the reader is familiar with Georgi's work. In order to study the
cyclunparticle analogue all we need is the general form of two-point
functions of cyclunparticle operators. For this, it is necessary to specify
the transformation properties under $SO(N)$ and the Lorentz group of the
operators in the two-point function. For example, if operators $\CO_1$ and
$\CO_2$ have scaling dimensions $\Delta_1$ and $\Delta_2$, respectively,
and  are scalars under $SO(N)$, then\footnote{Here and after we define
two-point functions in momentum space without the usual
$(2\pi)^4\delta^{(4)}(0)$ factor.} 
\eqn{ \vev{\CO_1(p)\CO_2(-p)} = C
(-p^2-i\epsilon)^{\frac12(\Delta_1+\Delta_2-4)},}
where $C$ is a constant. Here and below we write correlation functions in
Minkowski space. By contrast, for an $SO(N)$-vector of scalar operators,
$\CO_a$, with matrix of scaling dimensions $\Delta_{ab}$, one has
\eqn{ \vev{\CO_a(p)\CO_b(-p)}=(-p^2-i\epsilon)^{-2}
[(-p^2-i\epsilon)^{\frac12(\Delta+Q)}C(-p^2-i\epsilon)^{\frac12(\Delta-Q)}]_{ab},
}[twopointab]
where now $C$ is an $N\times N$ matrix of constants. Similarly, for an
$SO(N)$-vector of vector operators $\CO^\mu_a$,
\eqn{ \vev{\CO^\mu_a(p)\CO^\nu_b(-p)}=(-p^2-i\epsilon)^{-3}
[(-p^2-i\epsilon)^{\frac12(\Delta+Q)}(p^2g^{\mu\nu}C_1 + p^\mu p^\nu
C_2)(-p^2-i\epsilon)^{\frac12(\Delta-Q)}]_{ab} ,}[twopointvec]
where now both $C_1$ and $C_2$ are $N\times N$ matrices. Note that, as
opposed to the case of CFTs, $C_1$ and $C_2$ are independent. (In CFTs one
may simultaneously diagonalize $\Delta$, $C_1$ and $C_2$, and then give the
entries of $C_2$ in terms of the corresponding diagonal entries in $C_1$
and $\Delta$.) Similar expressions can be immediately written for other
$SO(N)$ representations and for correlators with mixed representations,
e.g.,
\eqn{
\vev{\CO_a(p)\CO^\prime(-p)}=(-p^2-i\epsilon)^{\frac12(\Delta^\prime-4)}
[(-p^2-i\epsilon)^{\frac12(\Delta+Q)}]_{ab}C_b.}[twopointmixed]

\newsec{Cyclunparticle phase space}
In his first paper Georgi considers rates $r(X\to Y {\cal U})$ for a state
$X$ with one or more standard particles to a state $Y$ of one or more
standard particles and one unparticle ${\cal U}$. If the state $X$
consists of a single particle, $r$ stands for a decay rate, else it stands
for a cross section. In either 
\eqn{ dr= \kappa(2\pi)^4\delta^{(4)}(P_X- p_Y-p_{\mathcal{U}}) \,d\Phi(Y)
\,d\Phi(\mathcal{U}) \,|\mathcal{M}|^2, }
where $P_X$ is the total 4-momentum of the initial state $X$,
$\mathcal{M}(X\to Y {\cal U})$ is the transition amplitude and $\kappa^{-1}
= 2\sqrt{P_X^2}$ or $4\sqrt{(P_{X_+}\cdot P_{X_-})^2-P_{X_+}^2P_{X_-}^2}$
depending on whether $r$ is a decay rate or a two-particle cross section
(here $X_+$ and $X_-$).  The phase space factor $d\Phi(Y)$ is the product
of single (normal) particle ones,
$(2\pi)\delta(p^2-m^2)\theta(p^0)\frac{d^4p}{(2\pi)^4}$ for a particle of
mass $m$. The novelty in unparticle physics is the phase space factors for
the unparticle. This can be obtained from the absorptive part of the
two-point function. For example, for a scalar operator (see Eq.~(4.4) in
Ref.~\rcite{Grinstein:2008qk}):
\eqn{
d\Phi(\mathcal{U})=C_{\mathcal{O}}(p^2)^{\Delta_{\mathcal{O}}-2}\theta(p^2)
\theta(p^0)\frac{d^4p}{(2\pi)^4},}
where $\Delta_{\mathcal{O}}$ is the scaling dimension of the CFT operator
$\mathcal{O}$.  The constant $C_{\mathcal{O}}$ depends on the normalization
of this operator. While this does not properly belong in the phase space,
it does belong in the product of phase space and amplitude-squared, so
nothing is lost by including it here rather than in $|\mathcal{M}|^2$. In
any case, the important point that Georgi makes is that the phase space for
the invisible `unparticle' is unusual in that it behaves much like a
multi-particle state with fractional particle number
$\Delta_{\mathcal{O}}$. This conclusion is independent of the specific
$X\to Y {\cal U}$ process considered.

The extension to cyclunparticles is immediate. One need only replace the
unparticle phase space, $d\Phi(\mathcal{U})$, with the cyclunparticle one,
$d\Phi(\mathcal{C})$. This can be determined from the absorptive part of
the two-point function, obtained by taking the imaginary part of, for
example, Eqs.~\eqref{twopointab}--\eqref{twopointmixed}. The phase space
for the cyclunparticle associated with some operator  $\CO$ is 
\eqn{d\Phi(\mathcal{C}_{\CO}) =
F(p^2)\theta(p^2)\theta(p^0)\frac{d^4p}{(2\pi)^4} }[cyclunPS] where
$F(p^2)$ is
the coefficient of $\theta(p^2)$ in the imaginary part of the two-point
function of~$\CO$.

Consider, for example, the case that $Y$ consists of a single particle of
mass $m$. Then the $\text{particle}+\text{cyclunparticle}$ production
differential rate is 
\eqn{ \frac{dr}{dE}= \frac{\kappa}{4\pi^2}|\mathcal{M}|^2\sqrt{E^2-m^2}\,
F(p^2), }
where $E=P_X\cdot p_Y/\sqrt{P_X^2}$ is the observable energy in the CM
frame and $p^2=P_X^2+m^2-2\sqrt{P_X^2}E$. At large CM energy (negligible
mass) this simplifies. In terms of the fraction of the total energy in the
cyclunparticle, $\xi=1-2E/\sqrt{P_X^2}$, we have
\eqn{ \frac{dr}{d\xi}= \frac{\kappa}{16\pi^2}|\mathcal{M}|^2\,P_X^2\,
(1-\xi)F(P_X^2\xi).  }
Our task is then to compute the function $F(p^2)$ for specific
cyclunparticles, at least for some specific  cases. 

Consider for definiteness an $SO(N)$-vector, scalar operator in the SIT,
$\CO_a(x)$, coupled to an external source $\chi$ through
$\mathcal{L}\supset g_a\chi\CO_a+\mbox{h.c.}$. This leads to the following
tree-level $\chi\rightarrow\chi$ forward scattering amplitude:
\eqn{
\mathcal{M}^{\text{fwd}}=g_ag_b|\chi|^2\left[(-p^2-i\epsilon)^{\frac12(\Delta+Q)-1}
C (-p^2-i\epsilon)^{\frac12(\Delta-Q)-1}\right]_{ab}\,.} 
Taking its imaginary part we obtain
\begin{multline}
  F(p^2)=-g_ag_b\left[(p^2)^{\frac12(\Delta+Q)-1}\left\{\cos\left[\left(\tfrac{\Delta+Q}{2}\right)\pi\right]
  C \sin\left[\left(\tfrac{\Delta-Q}{2}\right)\pi\right]\right.\right.\\
  \left.\left.+\sin\left[\left(\tfrac{\Delta+Q}{2}\right)\pi\right] C
  \cos\left[\left(\tfrac{\Delta-Q}{2}\right)\pi\right]\right\}(p^2)^{\frac12(\Delta-Q)-1}\right]_{ab}.
\end{multline}
This, with Eq.~\eqref{cyclunPS}, gives the phase space for the
cyclunparticle corresponding to the linear combination of operators
$g_a\CO_a$.

Similarly, for an $SO(N)$-vector, vector operator $\CO_a^\mu(x)$ in the
SIT, coupled to an external source $\chi_\mu$ through $\mathcal{L}\supset
g_a\chi_\mu\CO_a^\mu+\mbox{h.c.}$, the forward scattering
$\chi\rightarrow\chi$ amplitude is
\eqn{ \mathcal{M}^{\text{fwd}}=- g_a g_b
\left[(-p^2-i\epsilon)^{\frac12(\Delta+Q)-1} \left(\chi\cdot\chi^\dagger
C_1+\frac{|\chi\cdot p|^2}{p^2} C_2 \right)
(-p^2-i\epsilon)^{\frac12(\Delta-Q)-1}\right]_{ab}\,.}
We obtain the phase space factor $F(p^2)$ for the cyclunparticle
corresponding to the linear combination of operators
$g_a\epsilon_\mu\CO_a^\mu$ by taking the imaginary part of the forward
scattering amplitude:
\begin{multline} F(p^2)=g_ag_b\left[(p^2)^{\frac12(\Delta+Q)-1}
  \left\{\cos\left[\left(\tfrac{\Delta+Q}{2}\right)\pi\right]\widetilde{C}\sin\left[\left(\tfrac{\Delta-Q}{2}\right)\pi\right]\right.\right.\\
  \left.\left.+\sin\left[\left(\tfrac{\Delta+Q}{2}\right)\pi\right]\widetilde{C}\cos\left[\left(\tfrac{\Delta-Q}{2}\right)\pi\right]\right\}(p^2)^{\frac12(\Delta-Q)-1}\right]_{ab}\,,
\end{multline}
where
\eqn{ \widetilde{C}=\epsilon\cdot\epsilon^\dagger C_1+\frac{|\epsilon\cdot
p|^2}{p^2}C_2\,.}

\newsec{Cyclunparticle exchange}
The second class of examples studied by Georgi corresponds to processes in
which both initial and final states contain only standard particles, but
there are virtual unparticle contributions to the amplitude.  

Consider for definiteness an $SO(N)$-vector, scalar operator in the SIT,
$\CO_a(x)$, coupled to external sources $\chi^{(1)}$ and $\chi^{(2)}$
through $\mathcal{L}\supset
g_a^{(1)}\chi^{(1)}\CO_a+g_a^{(2)}\chi^{(2)}\CO_a+\mbox{h.c.}$. The sources
$\chi^{(1)}$ and $\chi^{(2)}$ model the standard particle initial and final
states and the coupling constants $g_a$ characterize the strength of the
interaction.  This leads to the following $s$-channel cyclunparticle
exchange contribution to the $\chi^{(1)}\rightarrow\chi^{(2)}$ scattering
amplitude,
\eqn{ \mathcal{M}^{\text{cyc}}
=g_a^{(1)}g_b^{(2)}\chi^{(1)}\chi^{(2)*}\left[(-p^2-i\epsilon)^{\frac12(\Delta+Q)-1}
C (-p^2-i\epsilon)^{\frac12(\Delta-Q)-1}\right]_{ab}\,.} 

Similarly, for an $SO(N)$-vector, vector operator $\CO_a^\mu(x)$ in the SIT,
coupled to external sources $\chi_\mu^{(1)}$ and $\chi_\mu^{(2)}$ through
$\mathcal{L}\supset
g_a^{(1)}\chi_\mu^{(1)}\CO_a^\mu+g_a^{(2)}\chi_\mu^{(2)}\CO_a^\mu+\mbox{h.c.}$,
the cyclunparticle contribution to the $s$-channel  scattering amplitude
for $\chi^{(1)}\rightarrow\chi^{(2)}$ amplitude is
\begin{multline*} \mathcal{M}^{\text{cyc}}=- g_a^{(1)}g_b^{(2)}
  \Bigg[(-p^2-i\epsilon)^{\frac12(\Delta+Q)-1}
  \Bigg(\chi^{(1)}\,\cdot\chi^{(2)*} C_1\\ +\frac{\chi^{(1)}\cdot p\,
  \chi^{(2)*}\cdot p}{p^2} C_2 \Bigg)
  (-p^2-i\epsilon)^{\frac12(\Delta-Q)-1}\Bigg]_{ab}\,.  \end{multline*} 

Consider, for example, an amplitude for which the SM gives an $s$-channel
contribution mediated by a photon, say, $e^+e^-\to\mu^+\mu^-$. If a vector
cyclunparticle couples to the same currents as the photon, then the
interference between the photon and cyclunparticle exchange gives a
fractional deviation of the cross section,
\eqn{ \frac{\sigma-\sigma_{\text{SM}}}{\sigma_{\text{SM}}}=2R+R^2+I^2,
}[cycExc]
where
\eqna{
R&=p^2\frac{g_a^{(1)}g_b^{(2)}}{e^2}\left[(p^2)^{\frac12(\Delta+Q)-1}\left\{\cos\left[\left(\tfrac{\Delta+Q}{2}\right)\pi\right]C_1\cos\left[\left(\tfrac{\Delta-Q}{2}\right)\pi\right]\right.\right.\\
&\quad\quad\quad\quad\quad\quad\quad\quad\left.\left.-\sin\left[\left(\tfrac{\Delta+Q}{2}\right)\pi\right]C_1\sin\left[\left(\tfrac{\Delta-Q}{2}\right)\pi\right]\right\}(p^2)^{\frac12(\Delta-Q)-1}\right]_{ab},\\
I&=-p^2\frac{g_a^{(1)}g_b^{(2)}}{e^2}\left[(p^2)^{\frac12(\Delta+Q)-1}\left\{\cos\left[\left(\tfrac{\Delta+Q}{2}\right)\pi\right]C_1\sin\left[\left(\tfrac{\Delta-Q}{2}\right)\pi\right]\right.\right.\\
&\quad\quad\quad\quad\quad\quad\quad\quad\left.\left.+\sin\left[\left(\tfrac{\Delta+Q}{2}\right)\pi\right]C_1\cos\left[\left(\tfrac{\Delta-Q}{2}\right)\pi\right]\right\}(p^2)^{\frac12(\Delta-Q)-1}\right]_{ab}\,.
}
It is straightforward to exhibit examples displaying corrections to
processes where other particles are exchanged. To better understand the
physical content of these expressions we turn to a more specific case.

\newsec{An $SO(2)$ example}
The formal expressions of the previous sections belie their complexity. To
better understand them we look at their explicit form in the simplest case,
that of $N=2$.  Consider first the matrix of vector correlators,
Eq.~\eqref{twopointab}. Using $Q_{12}=q$ and $\Delta= d\mathds{1} +
\gamma$, where $d$ is the naive dimension of the scalar operators\foot{We
have taken the two scalar operators to have the same naive dimensions. In
mass independent subtraction schemes operators with different naive
dimension do not mix with each other. While the cross correlators are
generically non-vanishing, one is not required to consider them, as would
be the case if they mixed under renormalization.} and $\gamma$ the matrix
of anomalous dimensions, we have
\eqna{ \vev{\CO_a(p)\CO_b(-p)}&=(-p^2)^{d-2+\frac12\Tr\gamma}\\ &
\quad\hspace{1cm}
\times\begin{pmatrix}c+\frac1{2\omega}(\gamma_{11}-\gamma_{22})s&
  \frac1{\omega}(\gamma_{12}+q)s\\ \frac1{\omega}(\gamma_{12}-q)s&
  c-\frac1{2\omega}(\gamma_{11}-\gamma_{22})s \end{pmatrix} \\
  &\quad\hspace{2cm} \times
  C\begin{pmatrix}c+\frac1{2\omega}(\gamma_{11}-\gamma_{22})s&
    \frac1{\omega}(\gamma_{12}-q)s\\ \frac1{\omega}(\gamma_{12}+q)s&
    c-\frac1{2\omega}(\gamma_{11}-\gamma_{22})s \end{pmatrix}
    ,}[twopointtwobytwo] 
where $c=\cos(\frac12\omega\ln(-p^2/\mu^2))$,
$s=\sin(\frac12\omega\ln(-p^2/\mu^2))$ and
$\omega=\sqrt{q^2-\frac14(\gamma_{11}-\gamma_{22})^2-\gamma_{12}^2}$.
Notice that the two-point function displays oscillations in momentum
provided $\omega$ is real, which in turn requires that $q^2>
\frac14(\gamma_{11}-\gamma_{22})^2+\gamma_{12}^2 $. Much as in the case of
unparticle physics, where large anomalous dimensions are assumed in the
strongly interacting CFT, for cyclunparticle physics we assume large $q$
and $\gamma$ with real $\omega$ in the strongly interacting SI model.

Taking the imaginary part of \eqref{twopointtwobytwo} we obtain the
function $F(p^2)$ in the phase space \eqref{cyclunPS}. For example, for the
cyclunparticle associated with the operator $g_1\CO_1$ the phase space
function $F$ is given by
\eqna{ F(p^2)&= g_1^2|p^2|^{\frac12\Tr\Delta -2}\bigg\{ \left[2c_d c_h s_h
(c + a  s) (s-a c) - s_d\left(c_h^2(c + a s)^2-s_h^2(s-a  c)^2  \right)
\right]C_{11} \\ &\quad+2\frac{\gamma_{12}+q}{\omega}\left[c_dc_h s_h\big(
s ( s -a c)- c( c+a s)\big)- s_d\big( c_h^2 s( c+a s)+s_h^2 c(s-a c)\big)
\right]C_{12} \\
&\quad-\frac{(\gamma_{12}+q)^2}{\omega^2}\left[2c_dc_hs_hcs +s_d(c_h^2
s^2-s_h^2 c^2)\right]C_{22}\bigg\}\,.  }[Fgiven]
Here $a=(\gamma_{11}-\gamma_{22})/2\omega$, $c_d=
\cos(\frac12\pi\Tr\Delta)$, $s_d= \sin(\frac12\pi\Tr\Delta)$, $c_h=
\cosh(\frac12\pi\omega)$, $s_h= \sinh(\frac12\pi\omega)$ and $c$ and $s$
are as before (modulo a minus sign), $c=\cos(\frac12\omega\ln(p^2/\mu^2))$
and $s=\sin(\frac12\omega\ln(p^2/\mu^2))$. 

\begin{figure}[ht] 
  \centering
  %\pstool*{phase-space}{\psfrag{p}[][]{$\hspace{12pt}\ln p^2$}
  %\psfrag{f}[][]{$F(p^2)$}}
  \includegraphics{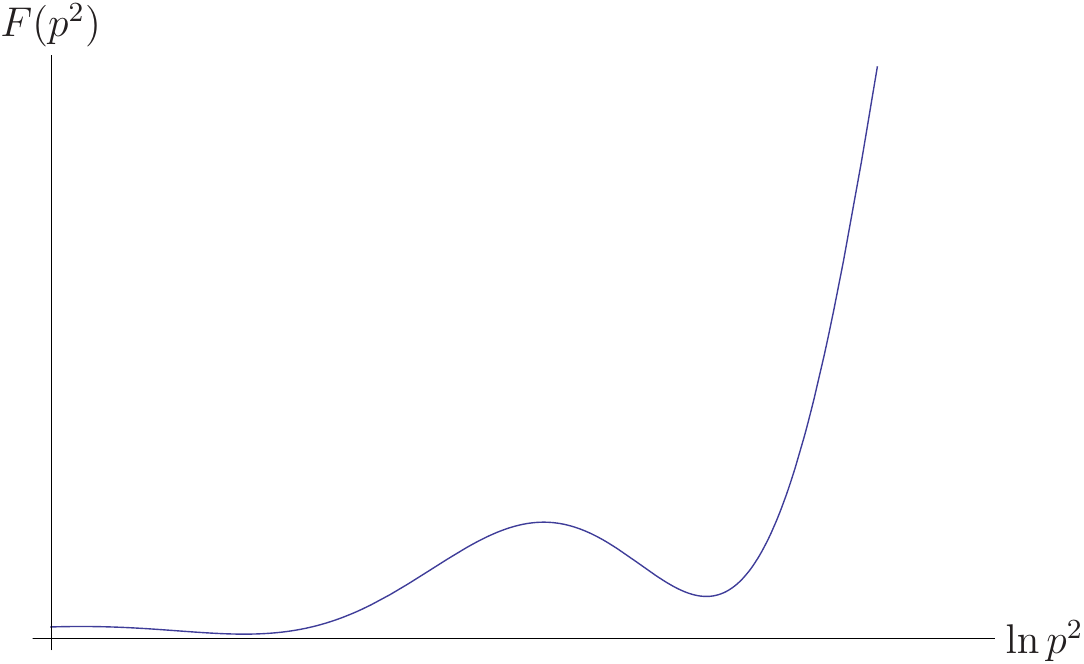}
  \caption{Plot of the function $F(p^2)$ determining the phase space for
  cyclunparticle production per Eq.~\eqref{cyclunPS}, for the case of a
  cyclunparticle that transforms as the first component of a vector of
  $SO(2)$.  The parameters used in this example are $d=2$,
  $\gamma_{11}=0.6$, $\gamma_{22}=0.4$, $\gamma_{12}=0.2$, $q=1.4$,
  $C_{11}=-2.9$, $C_{22}=-2.1$ and $C_{12}=0.2$; see Eq.~\eqref{Fgiven}. The
  frequency is $\omega=1.38$ and two periods are shown in the figure.
  }\label{fig:phaseSP}
\end{figure}

This function displays very unusual behavior. Beyond the scaling
corresponding to fractional particle number, familiar from unparticle
physics, the phase space exhibits oscillations with angular frequency
$\omega$. We show in Fig.~\ref{fig:phaseSP} the plot of $F$ for a specific
choice of parameters (as given in the figure caption).

A couple of important remarks: first, positivity of $F(p^2)$ is required by
unitarity. This, however, is not guaranteed by the form of the absorptive
part of the two-point function. Rather, positivity restricts the parameters
of the theory, namely the matrix of dimensions~$\Delta$, the coefficient of
the virial operator~$Q$ and the two-point function normalization
matrix~$C$. This is not unlike the situation in CFT---unitarity restricts
the representations of the conformal group precisely in that the dimensions
of operators are restricted \rcite{Mack:1977uf}. The solution to the
unitarity problem in SIT, namely, the conditions that~$\Delta$, $Q$ and~$C$
must satisfy to insure the positivity of the absorptive part of the
two-point function at all momenta, is not known even for the simple $N=2$
example of this section, let alone the general case.\foot{Except for
$SO(N)\times U(n)$ singlet operators, for which the condition on the
dimensions is given in Ref.~\rcite{Grinstein:2008qk}.} That the function
$F$ in the example of Fig.~\ref{fig:phaseSP} is positive only shows a
judicious choice of parameters, but the reader can easily find examples for
which $F$ fails to remain positive for all momenta.

Second, as pointed out above, oscillatory behavior in the two-point
function requires real $\omega$. This imposes conditions on the parameters
of the SIT that, however, are {\it a priori} independent from the unitarity
conditions.  But in a Lagrangian formulation of the theory the parameters
of the two-point function are derived from the coupling constants and hence
are not mutually independent. In that case the unitarity conditions must be
automatically satisfied (provided the Hamiltonian is Hermitian), but not so
the reality of $\omega$. This begs a question: is $\omega$ real for the few
known examples of perturbative SI RG-flows in $D=4-\epsilon$?  In the (few)
known examples the anomalous dimensions for the scalar fields already
receive contributions at one-loop order, while the equation for $Q$
requires that one goes to at least two-loops. Hence, $\omega$ is purely
imaginary in those cases. While there is no reason to forbid real $\omega$
in a strongly coupled SIT, this possibility is an assumption in this work.

\begin{figure}[ht]
  \centering
  %\pstool*{cross-section}{\psfrag{p}[][]{$\hspace{12pt}\ln p^2$}
  %\psfrag{s}[][]{$\ln|\Delta\sigma/\sigma_{\text{SM}}|$}}
  \includegraphics{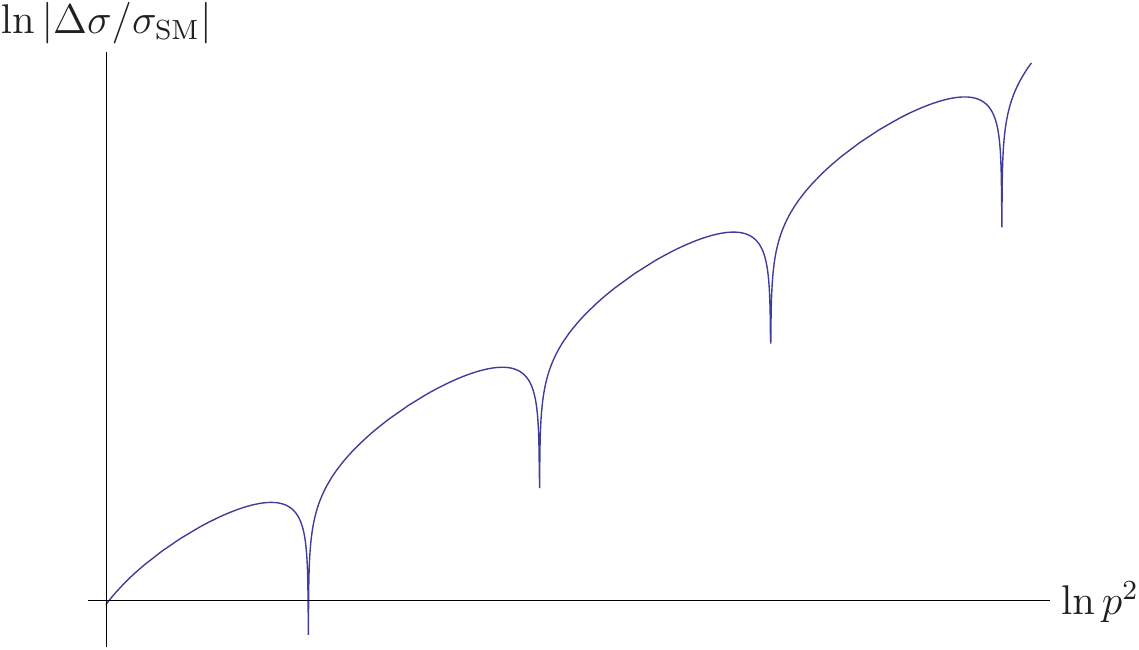}
  \caption{Plot of the fractional correction to the cross section, in
  arbitrary units, for $e^+e^-\to\mu^+\mu^-$ resulting from the exchange of
  a vector cyclunparticle transforming as the first component of a vector
  of $SO(2)$, as given per Eqs.~\eqref{crossSection} and \eqref{Qgiven}.
  The parameters used in this example are the same as in
  Fig.~\ref{fig:phaseSP}, except that $d=3$. Two periods are shown in the
  figure.}\label{fig:crossSec}
\end{figure}

We turn now to an example of a cyclunparticle exchange. Consider the
fractional deviation of the cross section for an $s$-channel photon
exchange, as in, say, $e^+e^-\to\mu^+\mu^-$. We have displayed the general
expression in Eq.~\eqref{cycExc}. For $N=2$ taking into account only the
interference term we can write more explicitly,
\eqn{ \frac{\sigma-\sigma_{\text{SM}}}{\sigma_{\text{SM}}} =
\frac{2g_1^{(1)}g_1^{(2)}}{e^2}(p^2)^{\frac12\Tr\Delta-1}G(p^2),
}[crossSection]
where for a cyclunparticle associated with the first component of the
$SO(N)$ vector only (that is, $g_a^{(i)}=0$ for $a\ne1$) we have
\eqn{\begin{aligned} G(p^2)&= \left[2s_d c_h s_h (c + a  s) (s-a c) +
  c_d\left(c_h^2(c + a s)^2-s_h^2(s-a  c)^2  \right) \right]C_{11} \\
  &\quad+2\frac{\gamma_{12}+q}{\omega}\left[s_dc_h s_h\big( s ( s -a c)- c(
  c+a s)\big)+ c_d\big( c_h^2 s( c+a s)+s_h^2 c(s-a c)\big) \right]C_{12}
  \\&\quad-\frac{(\gamma_{12}+q)^2}{\omega^2}\left[2s_dc_hs_hcs +c_d(c_h^2
  s^2-s_h^2 c^2)\right]C_{22}\,.  \end{aligned} }[Qgiven]
We show in Fig.~\ref{fig:crossSec} a plot of the fractional correction to
the cross section (assumed small so that the interference term is dominant)
for the same parameters as the example of Fig.~\ref{fig:phaseSP}.
The units are arbitrary since the normalization of the cyclunparticle
operator is free. Only a range of center of mass energy $\sqrt{p^2}/2$
corresponding to two cycles is shown. Oscillations are apparent. The
envelope of the correction grows with energy because the cyclunparticle
two-point function scales more slowly than the photon's.

\newsec{Conclusions}
We examined some consequences of coupling a non-conformal scale-invariant
sector to the standard model. Our most important result is that the
oscillating behavior of non-conformal scale-invariant correlation functions
is physical---it appears both in phase spaces and in amplitudes and cross
sections.  This leads to novel effects in standard model processes.

A simple example exhibiting oscillations has been provided.  Possibly more
interesting effects could be achieved by coupling $SO(3)$ cyclunparticles
to standard model flavor currents. Clearly, the potential model building
applications of non-conformal scale-invariant theories are largely unknown.

\ack{This work was supported in part by the US Department of Energy under
contract DOE-FG03-97ER40546, the National Science Foundation under Grant
No.\ 1066293 and the hospitality of the Aspen Center for Physics.}

\bibliography{cyclun_ref}

\end{document}